\documentstyle[12pt,aasms4]{article}
\def\lsim{~\rlap{$<$}{\lower 1.0ex\hbox{$\sim$}}}
\def\gsim{~\rlap{$>$}{\lower 1.0ex\hbox{$\sim$}}}

\begin{document}
\title{DEMONSTRATING DISCRETENESS AND COLLISION ERROR IN COSMOLOGICAL
N-BODY SIMULATIONS OF DARK MATTER GRAVITATIONAL CLUSTERING}
\author{Adrian L. Melott\altaffilmark{1}, Sergei F. Shandarin\altaffilmark{1},
Randall J. Splinter\altaffilmark{2,3} 
and Yasushi Suto\altaffilmark{4}}

\affil{melott@kusmos.phsx.ukans.edu, 
  sergei@kusmos.phsx.ukans.edu, randal@ccs.uky.edu, suto@phys.s.u-tokyo.ac.jp}

\begin{abstract}
Two-body scattering and other discreteness effects are unimportant 
in cosmological gravitational
clustering in most scenarios, since the dark matter has a small
particle mass. The collective field should determine evolution:
Two--body scattering in simulations violates the Poisson--Vlasov equations. 
We test this in PM, P$^3$M, Tree, and NGPM codes, noting
that a collisionless code
will preserve the one--dimensional character of plane wave collapse. We find
collisionality vanishing as 
the softening parameter approaches the mean
interparticle separation.              
Solutions for 
the problem are suggested, involving
greater computer power, PM--based nested grid codes, and a more
conservative approach to resolution claims. 

\altaffiltext{1}{Department of Physics and Astronomy, University of Kansas, 
Lawrence, KS 66045}
\altaffiltext{2}{Center for Computational Sciences, 325 McVey Hall, University of Kentucky, 
Lexington, KY 40506}
\altaffiltext{3}{Department of Physics and Astronomy, University of Kentucky, 
Lexington, KY 40506}
\altaffiltext{4}{Department of Physics and RESCEU, University of Tokyo, 
Tokyo 113, Japan}

\end{abstract}


\section{Introduction}

In the limit of small particle masses, a system of self--gravitating masses
is described by the Poisson--Vlasov equations: the particle--particle
scattering  
becomes unimportant, and the
evolution approaches that of a continuous system with a time--dependent
potential (Chandrasekhar 1942; Sellwood 1987).  N-body codes have a
small number of high mass particles compared to a Universe of
unclustered dark matter.  It is not clear whether the ensemble converges to some
or all of the properties of the right solution (Melott 1981). In fact, one may
conclude that if 
``...much of the mass in the universe comprises an invisible component
(the missing mass) there is no guarantee that the galaxies have ever
acted as point particles. If this were the case, the
results from N--body experiments would not apply to the real universe.''
(Hockney and Eastwood 1981, 1988).
The purpose of the Letter is to present preliminary results from a longer study
to warn of possible problems. 

The mean--field approach is typified by the 
particle--mesh, or PM method (Doroshkevich et. al 1980; Melott 1981, 1982b;
Klypin \& Shandarin 1983).  
Particles move in a 
gravitational potential computed on a 
mesh.                       
The shortcoming is that there are no valid results
below the mesh scale, since potential and density are smoothed over that
scale.  So far no errors have been reported other than this (rather
serious) 
limitation.

Short--range forces may be added to preserve the $r^{-2}$ force law in
close encounters. P$^3$M (Particle--Particle--Particle Mesh) 
(Hockney \& Eastwood 1981; Efstathiou \& Eastwood 1981)
and Tree codes (Suginohara et al. 1991) are two examples, although
more recently codes based on adaptive mesh refinement (Pen 1995;
Suisalu \& Saar 1995; Gelato et al. 1996; Kravtsov et al. 1996), have been used. 
Generally this approach improves resolution of the
Green function for the Poisson equation without improving
the resolution of the source term.  For this reason we call them HFLMR (High
Force Low Mass Resolution)  codes. Roughly isotropic contraction of
clumps is often used  to 
justify this approach.  Tests made by Kuhlman et al. (1996)
on generic smooth initial perturbations
do not support isotropic collapse. To prevent the formation of tight
binaries which slow down execution, all codes resort to
force softening, so that on scales less than $\epsilon$ the force law
is softer than $1/r^2$.  Values of $\epsilon$ (in units of the mean
interparticle separation $n^{-1/3}$) of 0.01 to 0.2
are common and results are usually presented down to $\epsilon$.  We
will show that $\epsilon \sim 1$ is  needed to maintain a
collisionless, quasi--continuous system.

Computer codes are often cross-checked for convergence, but a common
assumption may lead to a common error. Agreement with exact solutions
is better, but not easy.
One classic test for two--body scattering error is mass
segregation. Particles of higher mass settle to 
inner parts of bound systems due to equipartition
proceeding by two--body scattering. Efstathiou \& Eastwood (1981) found
strong segregation in P$^3$M.  This result appears to have been largely
ignored.  Peebles et al (1989) verified that it could be 
suppressed in PM with $\epsilon\sim 1$. In an
equal mass system, like most cosmological simulations, this error
may exist but not result in segregation.  Suisalu \& Saar (1996)
examined deflections and found an indication of trouble in a P$^3$M
code, but their original method was unable to show whether the scattering was due
to mean field or two-body fluctuations.

\section{Plane--Symmetric Collapse}

We suggest a new type of test (symmetry-breaking) for
codes in the nonlinear regime without an
exact solution.  We use a simple system with a clear
prediction: plane--wave collapse.  (Of course it could be spherical collapse,
two-dimensional collapse onto a filament, or any other type of
symmetric collapse.)  This has an exact solution up to shell crossing
(Zel'dovich 1970; Shandarin \& Zel'dovich 1989), and was used by Efstathiou
et al. (1985) in code testing.  However, they worked only in the precollapse
regime and along
coordinate axes so no collisions were possible.  Obviously a collisionless
system with only one--dimensional perturbations 
should remain one-dimensional. This is the basis of our test,
violation of which means the code is collisional or
otherwise erroneously scattering particle orbits.

We make the test more relevant by tilting the plane of
collapse relative to the simulation cube.
We set up a single perturbation wave $\bf k$ = (2,3,5)$k_f$
($|{\bf k}| = 6.16k_f$, where
$k_f$ is the fundamental mode) by Fourier transform on a grid of
$64^3$ particles. We began with an amplitude $\delta \equiv 
(\rho - {\bar \rho}) / {\bar \rho} \sim 0.1$, and
evolved for an expansion factor of 7.7 after the first shell crossing,
during which collisions can happen.  While the physical
system should have no scattering, near misses may generate them numerically.  The role of the symmetry is only to make it
detectable.
To perform the comparison we used a PM (Melott 1981, 1986), a P$^3$M (kindly
supplied by H. Couchman 1991) and a Tree-code (Suginohara et al. 1991).
We also tested the Nested--Grid Particle--Mesh (NGPM) code 
(Splinter 1996).
All runs had identical (publicly available) initial conditions.  The initial
conditions for the NGPM code were generated in the 
above manner for both the coarse and fine grid. We also did cross
check runs in which the perturbation ${\bf k}=(0,0,6)k_f$ was {\it not} tilted
with respect to the cube.                                 

The PM run was performed on a 64$^3$ mesh and duplicated
on a 128$^3$ mesh to emulate a sometimes used modification as well as verify
the code-independence of our results. 
PM tests were done with traditional two-point differencing and the Melott
(1986) improved force resolution staggered mesh scheme. There was no significant difference in scattering
and we report the latter here.
We performed otherwise identical P$^3$M and Tree tolerance 
parameter $\theta=0.2$ runs with $\epsilon=$ 
0.1, and 1.0, plus a transitional P$^3$M run with $\epsilon=$ 0.5.  
In the P$^3$M code, we used two choices of time integration variable
and varied the timestep greatly, assuring satisfaction of both Courant and
leapfrog stability conditions.  The PM and NGPM codes automatically test and
adjust timesteps as needed.
The adaptive smoothing length capability of
the P$^3$M code was turned off as suggested by Gelb and Bertschinger
(1994). The NGPM code
had a refinement factor of 8, putting it close in spatial
resolution to the $\epsilon = 0.1$ P$^3$M run, but with 512 times increased mass
resolution (an ``HFHMR'' code).
Results of a much more extended study will 
be presented elsewhere.            

Figure 1 shows the overall configuration of the PM system after collapse. All
runs  
look roughly similar.  Differences between tilted runs  are shown in 
Figure 2, in which slices of one collapsed planar region are 
projected along initial perturbation axis. The only
inhomogeneity should be projection of the initial lattice onto this plane.
Some runs show clumping, 
suggesting scattering error.
What all the erroneous HFLMR runs (the P$^3$M and Tree
code runs with $\epsilon< 1$, and the 128$^3$ mesh PM run) share are softening
lengths shorter than the mean interparticle separation.
The runs that performed well (normal PM, P$^3$M and Tree with $\epsilon=$ 1,
and NGPM) all have softening comparable to this distance; of course
for NGPM this is a considerably smaller distance, but at no collision
penalty.
(Axis--aligned PM and P$^3$M runs show the lattice, with no clumping visible.)

We use as one quantitative measure the distribution of particle
velocities.  They should be strictly normal to the planes; we separate
them into components along the normal and in the plane $V_{plane}
= \sqrt{V_{p1}^2+V_{p2}^2}$.  
Figure 3
shows scatter plots for 1000 randomly selected particles from each of
our runs. Many are
hidden by superposition.  The correct result is a line along the
$V_{norm}$ axis. This is approached only by non--sparse PM and NGPM, by P$^3$M
and Tree as the short--range force is turned off, and by axis--aligned
runs which have only head--on collisions.
With $\epsilon =0.1$, the most
common choice, the error is large.  

The relative error can be made quantitative by comparing 
the median speed  in the plane to the median speed along the normal,
as shown in Table 1.
Another
measure is the kinetic energy;  the mean from motion in the plane
and along the normal are also shown in Table 1. Lastly, we show the median
value of $d_{plane}$, the distance in mesh units particles have strayed off the 
normal trajectory. All values are the mean or median of 10,000
particles (subgrid particles in NGPM).
Our axis--aligned PM and P$^3$M runs had {\it zero}
off--normal velocity (within computer precision).

Figure 4 shows a phase--space diagram of a single sheet, including the
normal displacement and velocity, with the other four phase space
dimensions suppressed.  The correct solution is a well--known spiral
(Doroshkevich et al. 1980; Melott 1982a; Bond et al. 1983).
The codes that preserve this pattern are those with
softening comparable to the mean interparticle separation.

We can verify that scattering is  from encounters and not the initial
gravity fields by noting that off-normal components are small until
shell crossing in all codes; they increase strongly in the inclined
HFLMR codes as particles pass each other.

\section{Discussion}

We have shown that HFLMR computational methods in widespread use for 
gravitational clustering in cosmology perform incorrectly on a simple
test problem, as 
a consequence of trying to model a continuous system with
discrete masses. The PM and NGPM methods (as normally used) are able to handle this test
because there is no evasion of the discreteness limitation.
PM can be forced to fail by increasing the lattice resolution beyond
appropriate limits. HFLMR methods work properly if the short range
force is turned off or if they are forced to alignment with the coordinate
axes.

As convergence
to the proper behavior is very slow (e.g. Hockney, 1971), past comparisons by varying
particle number have not revealed this (e.g. Efstathiou \& Eastwood 1981).
Coupling these
incorrectly evolved systems to hydrodynamics will guarantee that
it is being done in the wrong background gravitational
potential.  We do not claim the effect will move to larger scales. Melott and Shandarin
(1990), Little et
al. (1991), and Melott \& Shandarin (1993) have shown that small
scale effects scarcely propagate to large scales, but more quantitative
study is needed.  However, errors
would only stop growing in voids or in regions where the particle
density exceeds $\epsilon^{-3}$.

Questions may be raised about the relevance of our example. Galaxies
are not infinite planes. However, the first collapse on any scale is
expected to be sheet--like (Shandarin et al. 1995; Kuhlman, et
al. 1996; Gouda 1996) so there is ample opportunity for this 
situation to arise.
Furthermore, collisionality operates in the absence of symmetry; our planar
collapse study
simply makes it starkly obvious.
One may argue that since collapsed pancakes are unstable to 
small--scale perturbations, the HFLMR codes
model this correctly, justifying
the results they give for small $\epsilon$. 
Since there is no small-scale power in the initial conditions,
these codes are artificially producing
power on small--scales by the growth of shot noise. The results of a simulation
should be a consequence of initial conditions that were imposed. 
This is illustrated in the orientation-dependence of the 
HFLMR codes: Since we get two completely different results depending
on orientation, one must ask, ``Which is correct?''
Most importantly, this serves to raise the question of whether a code 
performs well overall in a complex nonlinear problem when it cannot replicate a
simple test case. 
As this Letter was going to press, we learned of Park (1997), in which
spherical collapse is studied, producing conclusions close to ours.
Values $\epsilon = 0.01$ or even smaller are used in clustering studies.

One might hope that realistic cosmological scenarios with power on all
scales avoid this problem. Impressed perturbations might
overwhelm discreteness if the spectrum is normalized to the shot noise
level at the particle Nyquist frequency (Efstathiou et al. 1985). We
tested this by putting in an inclined plane wave close to the particle Nyquist
frequency, at the white noise amplitude. Again we
found strong scattering in a $\epsilon =0.1$ P$^3$M run, and
essentially none in PM. 
At this short wavelength the resolution
limitations of PM show themselves in its lower velocity
dispersion, so both codes are performing badly.
The accuracy of cosmological results from HFLMR codes remains an open question.

Suto (1991)  examined the  divergence  of  particle trajectories  in a
series of  N--body cosmological   simulations varying $\epsilon$.   He
found that the comoving trajectories diverged with time as $e^{\lambda
t}$, fitting $\lambda \simeq 0.05
\sqrt{G \bar{\rho} } \epsilon^{-1.2}$.
By requiring $e^{\lambda t_H} < \epsilon$,
where $t_H$ is  the Hubble time we require that deflection of
nearby   trajectories  by shot noise is  small.  Enforcing  such  a  condition
gives 
$\epsilon \gsim 1$, similar to our results, the results of Peebles et al.
(1989) and Suisalu and Saar (1996).

One promising method to achieve better force and mass resolution while
doing correct physics is nested--grid methods, which put more
particles in a region of interest. Such methods are shown here to
greatly reduce collisions and may allow the study of small--scale structure to
progress (e.g. Villumsen 1988; Anninos 
et al. 1994; Splinter 1996). 
Putting in a higher particle density
acknowledges the inability get something for nothing by sidestepping
the laws of physics.

ALM and SFS wish to acknowledge the financial support of the
NSF--EPSCoR program, NASA grant NAGW-3832, the National Center for
Supercomputing Applications, 
and Enn Saar for helpful comments.
RJS wishes to thank the Center for Computational
Sciences at the University of Kentucky and the Kentucky Space Grant
Consortium for financial support, and Hugh Couchman
for advice in the use of his P$^3$M code. YS thanks RESCEU (Research
Center for the Early Universe, University of Tokyo), and KEK (National
Laboratory for High Energy Physics, Japan).

\clearpage
\begin{figure}
\caption{The configuration of particles at the end of our PM simulation.
The other simulations look much the same except for more inhomogeneity
in some cases.
\label{fig:angles}}
\end{figure}
%
\begin{figure}
\caption{A slice of one of the planes from each cube, seen projected along the 
normal to the plane. The dimensions of the slice are $16 \times 16 \times 4$.
To construct the NGPM slice a slice of size $4 \times 4 \times 1$ was 
extracted from
the sub--grid particles and repeated periodically to produce a slice of size
$16 \times 16 \times 4$. This slice was then sampled to reduce the number of 
particles to roughly that of the other runs.
The arrangement of panels is: Top row--PM with one particle per cell,
PM with one particle per 8 cells (a common `resolution increasing'
procedure) and NGPM (Sub--grid).  Middle row--P$^3$M with
specified value of $\epsilon$.
Bottom row--Tree code with the specified $\epsilon$, and
the correct result, which was constructed by propogating particles
along normals to the plane; the appearance of lines is a tilted
projection of the cubic lattice.
This projection represents the standard for all except NGPM,
which shows correct appearance.
\label{fig:orthoslices}}
\end{figure}
%
\begin{figure}
\caption{Scatter plots of the absolute value of velocity components for 1000
particles randomly selected from each of the simulations, projected along the
normal to the plane of collapse, and in that plane. For correct physical
modeling, all points should lie along the x axis.  Each plot contains the
same number of points; many are superimposed.
Same arrangement as Fig. 2. Velocity units are Hubble velocity across one cell.
\label{fig:scatter}}
\end{figure}
%
\begin{figure}
\caption {Scatter plots of the normal component of the velocity for all
particles collapsed toward one of the pancake planes, against their 
displacement from the midpoint. The known solution is a sprial pattern,
whose development varies with resolution in some codes, but is totally
disrupted in others.
Same arrangement and units as in Fig. 3.
\label{fig:phase}}
\end{figure}
\newpage

\begin{deluxetable}{lcccccc}
\tablenum{1}
\tablewidth{0pt}
\tablecaption{Code Comparison--Plane Wave Test \label{table1} }
\tablehead{
\colhead{Code}              & 
\colhead{$med(V_{norm})$}   & \colhead{$med(V_{plane})$}   &
\colhead{${\bar T}_{norm}$} & \colhead{${\bar T}_{plane}$} &
\colhead{$med(d_{plane})$} } 
\startdata
PM                                 &  \phn  0.75 & 0.03 & 1.82 &      0.001     & 0.01  \nl
PM(N$_c=128^3$,N$_p=64^3$)         &  \phn  0.91 & 0.44 & 2.53 & \phn 0.22 \phn & 0.80  \nl
Sub-Grid(R=8)                      &  \phn  0.77 & 0.05 & 1.84 &      0.02    
& 0.003 \nl 
P$^3$M(${\epsilon=1.0}$)           &  \phn  0.70 & 0.05 & 1.99 &      0.004     & 0.03  \nl
P$^3$M(${\epsilon=0.5}$)         &  \phn  0.82 & 0.27 & 2.00 &     0.15 \phn
& 0.12  \nl 
P$^3$M(${\epsilon=0.1}$)           &  \phn  0.78 & 0.62 & 2.10 & \phn 0.76 \phn & 0.53  \nl
Tree(${\epsilon=1.0, \theta=0.2}$) &  \phn  0.57 & 0.01 & 1.82 &      0.0003     
& 0.02  \nl 
Tree(${\epsilon=0.1, \theta=0.2}$) &  \phn  0.81 & 0.62 & 2.08 &      0.79    
  & 0.45 \nl
\enddata

\end{deluxetable}
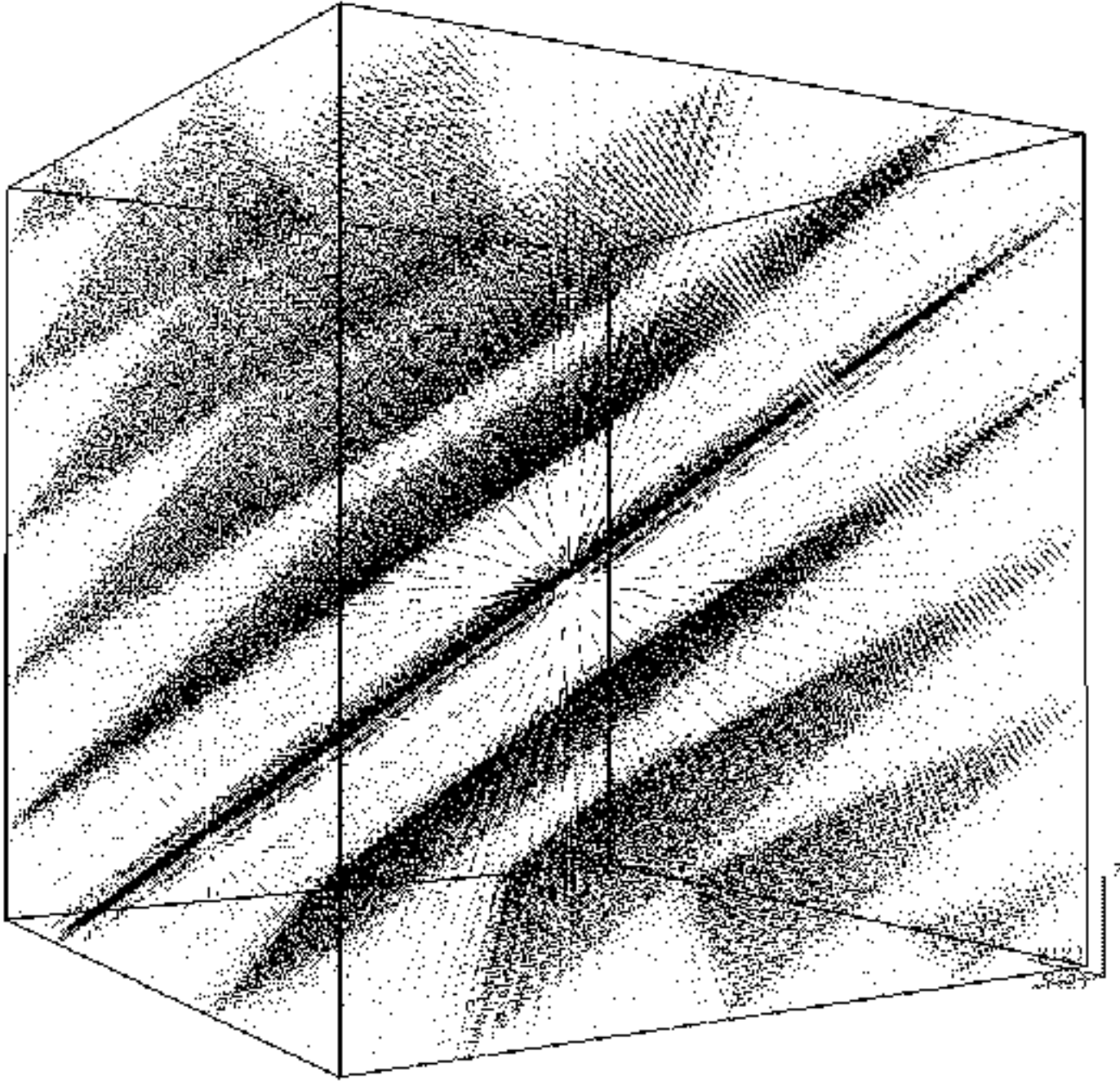
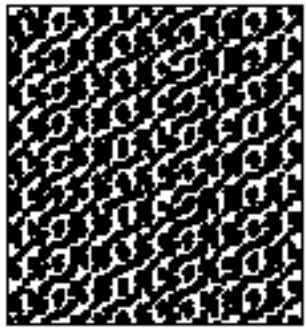
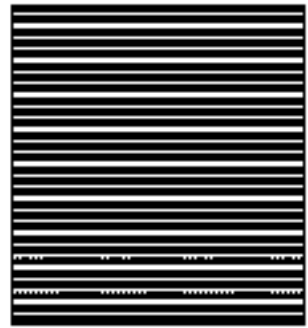
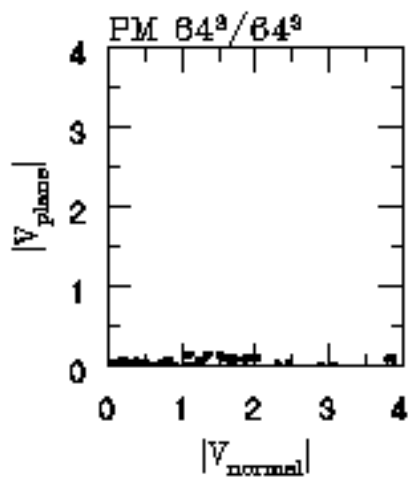
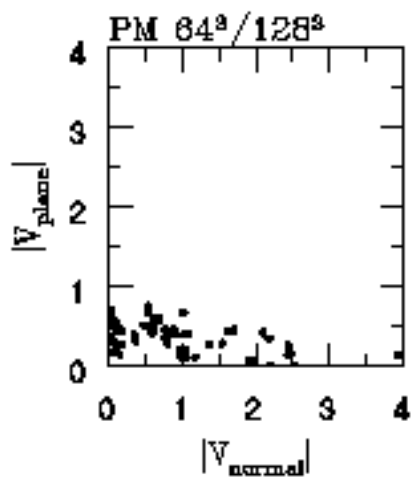
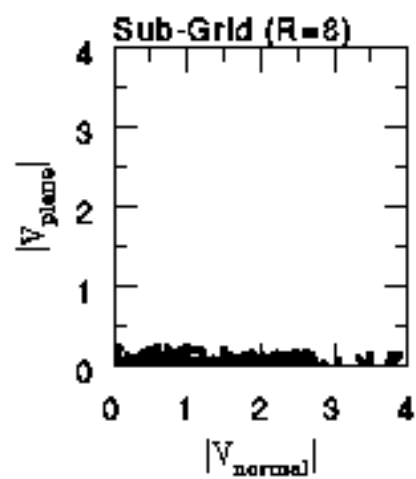
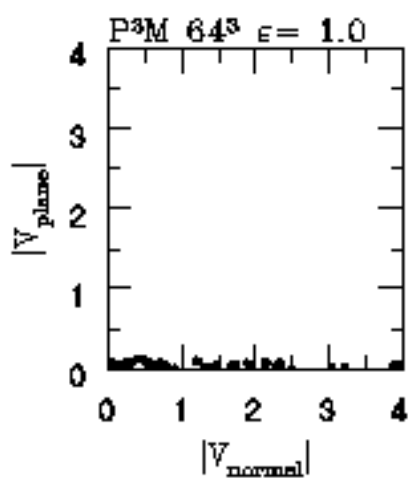
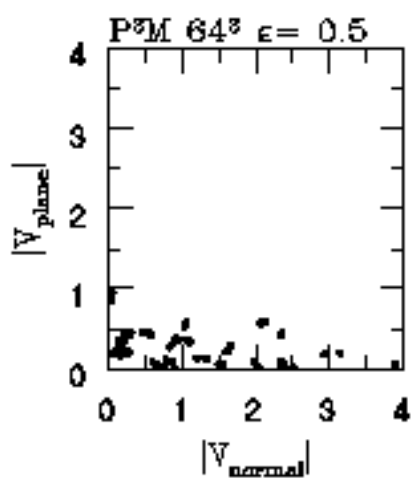
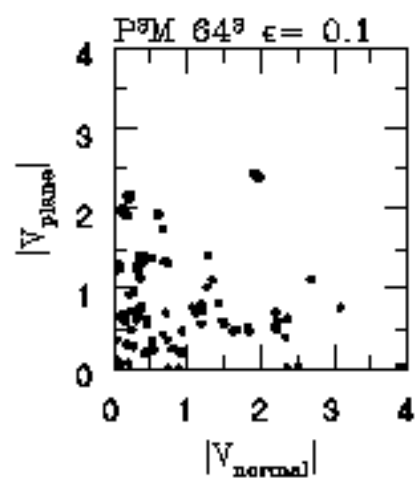
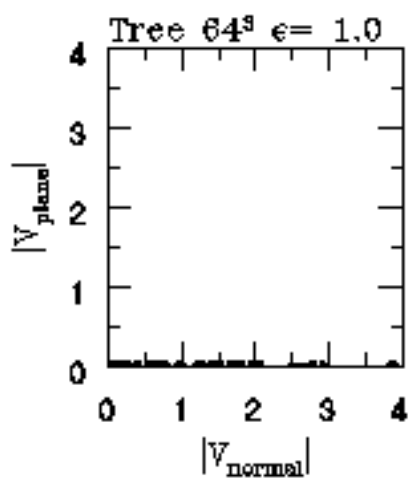
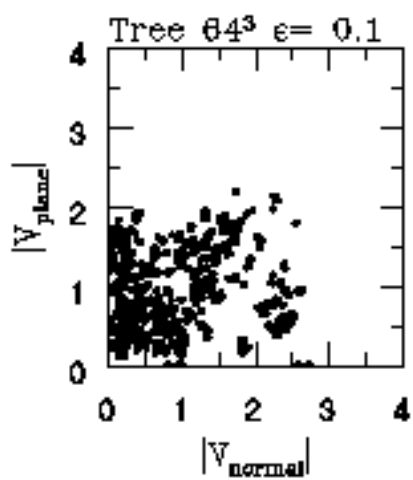
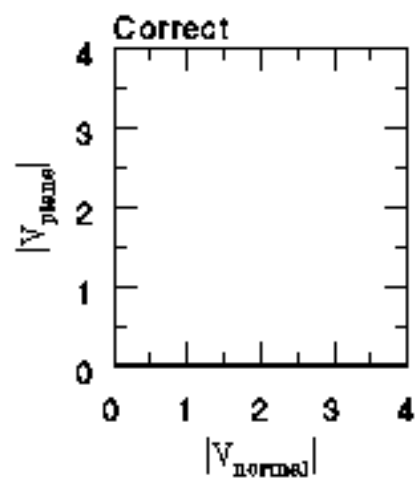
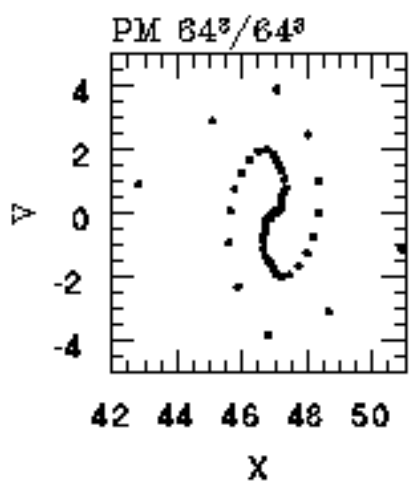
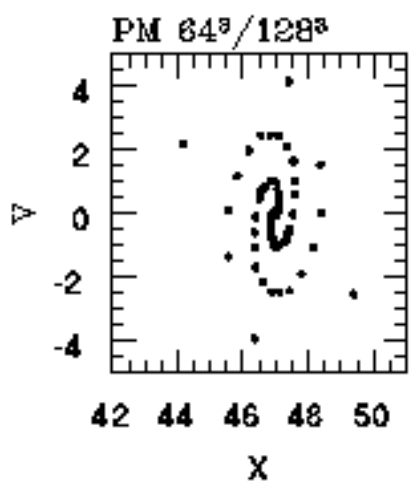
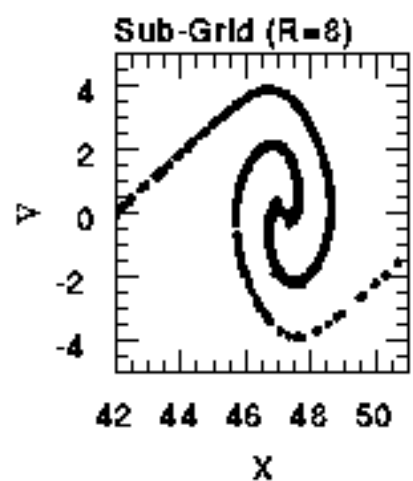
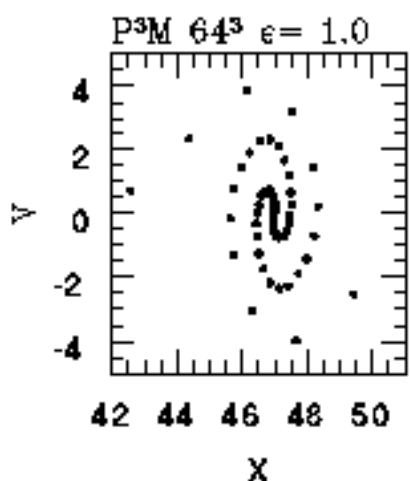
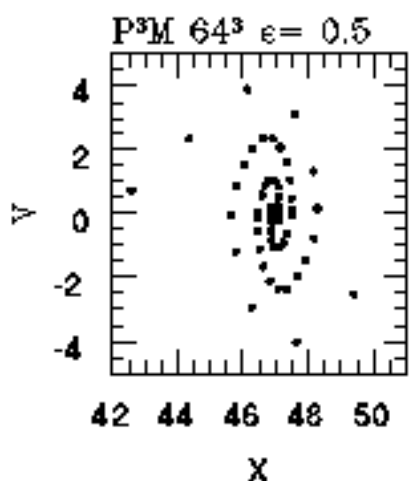
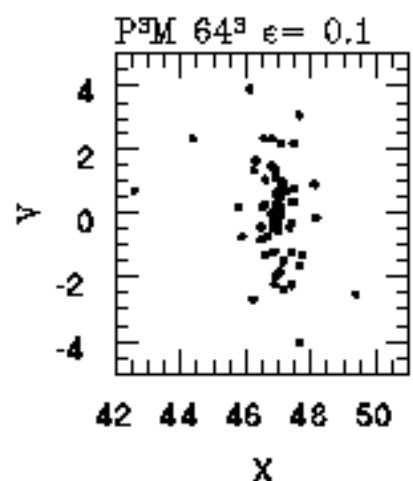
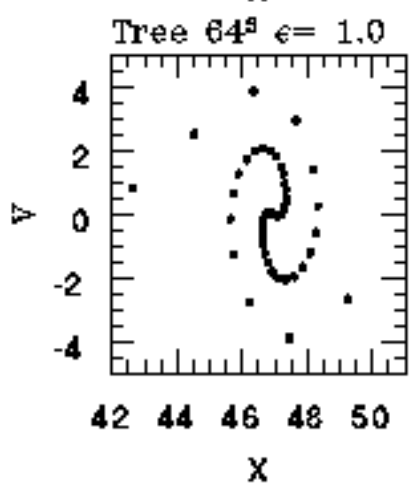
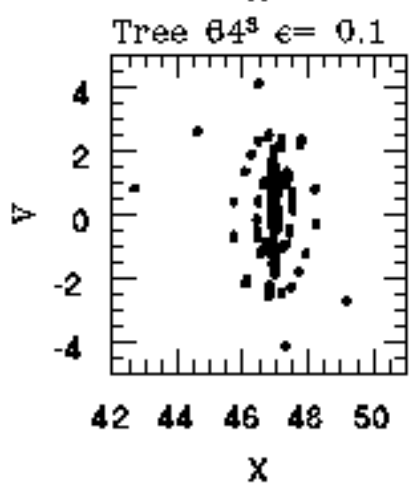
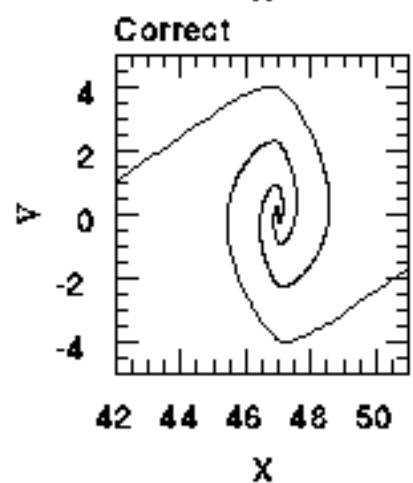
\end{document}